\documentclass[a4paper,11pt]{article}

\usepackage{amsthm}
\usepackage{mathptmx}
\usepackage{amssymb}
\usepackage{amsfonts}
\usepackage{tikz}
\usepackage{authblk}
\usepackage{amsmath,amscd}
\usepackage{pgfplots}
\usepackage{tikz}
\usepackage{amscd}
\usepackage{authblk}
\usepackage{geometry}
\usepackage{amsmath,amsthm,wasysym,txfonts,indentfirst,fancyhdr,amscd, multirow}

\newtheorem{defn}{Definition}[section]

\newtheorem{prop}{Proposition}[section]

\newcommand{\Z}{\mathbb{Z}}

\newcommand{\I}{\mathcal{I}}
\newcommand{\It}{\widetilde{\mathcal{I}}}
\newcommand{\Tt}{\widetilde{\mathcal{T}}}
\newcommand{\Dt}{\widetilde{\mathcal{D}}}
\newcommand{\Gt}{\widetilde{\mathcal{G}}}

\newcommand{\bfx}{\boldsymbol{x}}

\newcommand{\bfe}{\boldsymbol{e}}

\newcommand{\bfv}{\boldsymbol{v}}

\newcommand{\R}{\mathbb{R}}

\newcommand{\C}{\mathcal{C}}
\newcommand{\z}{\mathcal{Z}}
\newcommand{\T}{\mathcal{T}}
\newcommand{\D}{\mathcal{D}}
\newcommand{\La}{\mathcal{L}}
\newcommand{\Po}{\mathcal{P}}
\newcommand{\G}{\mathcal{G}}

\newcommand{\h}{\mathcal{H}}

\title{A group theoretical approach to structural transitions of icosahedral quasicrystals and point arrays}
\date{}
\author[1,4]{Emilio Zappa \thanks{ Current address: Courant Institute of Mathematical Sciences, New York University, NY. Electronic address: zappa@cims.nyu.edu.}}
\author[1,2,4]{Eric C. Dykeman}
\author[2,3,4]{James A. Geraets \thanks{ Current address: Institute of Biotechnology, University of Helsinki, Finland.}}
\author[1,2,4]{Reidun Twarock}
\affil[1]{Department of Mathematics,}
\affil[2]{Department of Biology,}
\affil[3]{Department of Physics,}
\affil[4]{York Center for Complex Systems Analysis, University of York, UK} 

\begin{document}
\maketitle

\begin{abstract}

In this paper we describe a group theoretical approach to the study of structural transitions of icosahedral quasicrystals and point arrays. We apply the concept of Schur rotations, originally proposed by Kramer, to the case of aperiodic structures with icosahedral symmetry;  
these rotations induce a rotation of the physical and orthogonal spaces invariant under the icosahedral group, and hence, via the cut-and-project method, a continuous transformation of the corresponding model sets. We prove that this approach allows for a characterisation of such transitions in a purely group theoretical framework, and provide explicit computations and specific examples. Moreover, we prove that this approach can be used in the case of finite point sets with icosahedral symmetry,  which have a wide range of applications in carbon chemistry (fullerenes) and biology (viral capsids).  

\end{abstract}

\section{Introduction}

Non-crystallographic symmetries appear in a wide range of physical and biological structures. Prominent examples are quasicrystals, alloys with long-range order lacking translational periodicity, discovered by Shechtman in 1984 \cite{quasicrystals}, fullerenes, molecules of carbon atoms arranged to form icosahedral cages \cite{kroto1}, and viral capsids, the protein containers which encapsulate the viral genomic material \cite{casjens}. Quasicrystals are described mathematically via cut-and-project schemes and model sets \cite{senechal, grimm, moody}, which correspond to infinite Delone point sets \cite{grimm}. On the other hand, the arrangement of proteins in a viral capsid or the atoms of fullerenes are modeled via finite point sets (arrays) with icosahedral symmetry. In particular, Caspar-Klug theory and generalisations thereof \cite{twarock} predict the locations and relative orientations of the capsid proteins via icosahedral surface lattices. Moreover, the three-dimensional organisation of viral capsids can be modeled via finite nested point sets displaying icosahedral symmetry at each radial level \cite{keef, jess, dechant2, zappa2}  or 3D icosahedral tilings \cite{david}. Such point sets can be characterised via affine extensions of the icosahedral group \cite{keef, jess}, the Kac-Moody algebra formalism \cite{dechant2}, or in a finite group theoretical framework \cite{zappa2}. The latter provides a link between the construction of quasicrystals and point arrays, the common thread being the orthogonal projection of points of a higher dimensional lattice into a space invariant under the icosahedral group. 

Quasicrystals undergo structural transitions as a consequence of changes of thermodynamical parameters. Experimental observations showed that aperiodic structures transform into higher order structures (crystal lattices) or other quasicrystalline solids \cite{steurer2, tsai, koster}. These transitions are characterised by the occurrence of a symmetry breaking, thus allowing for their analysis in the framework of the phenomenological Landau theory for phase transitions \cite{sethna, landau}. Similar transformations occur in virology; specifically, viruses undergo conformational changes as part of their maturation process, important for becoming infective, resulting in an expansion of the capsid \cite{erav}. Experiments and theoretical analysis \cite{zappa} proved that the initial and final states of the transformation retain icosahedral symmetry; however, little is known about the possible transition paths, which according to a first mathematical approach are likely to be asymmetrical \cite{paolo2}. 

In this paper we characterise structural transitions of icosahedral quasicrystals and point arrays in a group theoretical framework. The starting point is the crystallographic embedding of the icosahedral group into the point group of the hypercubic lattices in $\mathbb{R}^6$ \cite{levitov, zappa1, senechal}, which is a standard way of defining icosahedral quasicrystals with the cut-and-project method. We then construct continuous transformations between icosahedral structures based on the Schur rotation method, originally introduced by Kramer \emph{et al.} \cite{kramer, kramer2}, for the case of transitions between cubic and aperiodic order with tetrahedral symmetry. Specifically, we consider two distinct crystallographic representations of the icosahedral group sharing a common maximal subgroup $\G$, and define rotations in $SO(6)$ that keep the $\G$-symmetry preserved. Such rotations induce, in projection, continuous $\G$-symmetry preserving transformations of the corresponding model sets or point arrays. We show that these rotations are parameterised by angles belonging to a $k$-dimensional torus $\mathbb{T}^k$, which are candidates to be the order parameters of the Landau theory. Moreover, we classify, based on the results of our previous paper \cite{zappa1}, the possible boundary conditions of the transitions, which, in this context, correspond to specific angles in $\mathbb{T}^k$.

Structural transitions of quasicrystals and point arrays have been analysed previously with different approaches, for examples with the phason strain approach \cite{cheng}, which was later proved to be equivalent to the Schur rotation method \cite{torres}. Moreover, Indelicato \emph{et al.} \cite{paolo, giuliana} define higher dimensional generalisations of the Bain strain between crystal lattices which induce, in projection, structural transformations of quasicrystals. In the latter reference it is proved that the Schur rotation method is related to the Bain strain in the case of transitions between cubic to icosahedral order \cite{paolo}, and some specific examples are given in the case of transitions between icosahedral tilings of the space. In this work we demonstrate analytically that the Schur rotation is equivalent to a Bain strain transformation between two congruent six-dimensional lattices. We argue that the advantage of the approach discussed here is the possibilty of classifying the boundary conditions and identifying the order parameters of the transition, while the Bain strain allows only a parameterization of the transitions in terms of paths in the centralisers of the maximal subgroups. Therefore, this work paves the way for a dynamical analysis of the possible transition pathways, by constructing Hamiltonians invariant under the symmetry group of the system in terms of the order parameters of the transition and analysing the resulting energy landscapes, as in previous works \cite{zappa}.  

The paper is organised as follows. After reviewing, in Section \ref{embedding}, the crystallographic embedding of the icosahedral group, based on \cite{zappa1}, we provide, in Section \ref{schur}, the theoretical framework for the analysis of transitions of quasicrystals and point arrays with icosahedral symmetry, by applying the Schur rotation method and generalising it for any maximal subgroup $\G$ of the icosahedral group $\I$ . In Section \ref{schur_comp} we explicitly compute the Schur rotations for the three maximal subgroups of $\I$, namely the tetrahedral group $\T$ and the dihedral groups $\D_{10}$ and $\D_6$, classifying the possible boundary conditions of the transitions, and provide specific examples. We conclude in Section \ref{conclusions} by discussing further applications and future work. 

\section{Crystallographic embedding of the icosahedral group}\label{embedding}

Let $\La$ be a lattice in $\R^d$ with generator matrix $B \in GL(d,\R)$. The point group $\Po$ of $\La$ is the set of all the orthogonal transformations that leave $\La$ invariant:
\begin{equation*}
\Po := \{ Q \in O(d) : \exists M \in GL(d,\Z): QB = BM \}\,.
\end{equation*}
$\Po$ is finite and does not depend on the matrix $B$ \cite{zanzotto}. A (finite) group of isometries $G \subseteq O(k)$ is said to be \emph{crystallographic} in $\R^k$ if it leaves a lattice $\La$ in $\R^k$ invariant, or, equivalently, if it is the subgroup of a point group $\Po \subseteq O(k)$. The crystallographic restriction states that, for $k =2,3$, a crystallographic group $G$ must contain elements of order $1,2,3,4$ or $6$ \cite{grimm}. Following Levitov and Rhyner \cite{levitov}, we introduce the following:

\begin{defn}\label{cryst_def}
Let $G \subseteq O(k)$ be a finite non-crystallographic group of isometries. A \emph{crystallographic representation} of $G$ is a matrix group $\widetilde{G}$ satisfying the following conditions:
\begin{enumerate}
\item[\emph{($\mathbf{C1}$)}] $\widetilde{G}$ stabilises a lattice $\La$ in $\R^d$, with $d > k$, i.e. $\widetilde{G}$ is a subgroup of the point group $\Po$ of $\La$;
\item[\emph{($\mathbf{C2}$)}] $\widetilde{G}$ is reducible in $GL(d,\R)$ and contains an irreducible representation $\rho_k$ of $G$ of degree $k$, i.e. 
\begin{equation}\label{cryst}  
\widetilde{G} \simeq \rho_k \oplus \rho', \qquad \deg(\rho') = d-k.
\end{equation}
\end{enumerate} 
\end{defn}

The condition ($\mathbf{C2}$) is necessary for the construction of quasicrystals with $G$-symmetry via the cut-and-project method \cite{grimm, senechal}. 

Chiral icosahedral symmetry is described by the icosahedral group $\I$, which corresponds to the set of all the rotations that leave an icosahedron invariant. It has order $60$ and is the largest finite subgroup of the special orthogonal group $SO(3)$ \cite{artin}.  It is isomorphic to the alternating group $\mathfrak{A}_5$, and has presentation
\begin{equation*}
\I = \langle g_2, g_3 : g_2^2=g_3^3=(g_2g_3)^5=e \rangle, 
\end{equation*}
where $g_2$ and $g_3$ represent, geometrically, a two- and a three-fold rotation, respectively. The element $g_5:=g_2g_3$ is a five-fold rotation, hence $\I$ is non-crystallographic in $\R^3$. Its character table is the following ($\tau := \frac{1}{2} \left(1+\sqrt{5}\right)$ denotes the golden ratio, and $\tau' := 1-\tau$ its Galois conjugate):  
\begin{center}
{\small{
\begin{tabular}{l|c c c c c}
Irrep & $E$ & $\C(g_5)$ & $\C(g_5^2)$ & $\C(g_2)$ & $\C(g_3)$ \\
\hline
$A$ & 1 & 1 & 1 & 1 & 1 \\
$T_1$ & 3 & $\tau$ & $\tau'$ & -1 & 0 \\
$T_2$ & 3 & $\tau'$ & $\tau$ & -1 & 0 \\
$G$ & 4 & -1 & -1 & 0 & 1 \\
$H$ & 5 & 0 & 0 & 1 & -1 \\
\end{tabular}}}
\end{center}
The minimal crystallographic dimension of $\I$ is six (cf. Definition \ref{cryst_def}) \cite{levitov}; an explicit crystallographic representation $\It$ of $\I$ is given by \cite{zappa1}:
{\small{
\begin{equation}\label{gen}
\widetilde{\I}(g_2) =  \left( \begin{array}{cccccc}
0 & 0 & 0 & 0 & 0 & 1 \\
0 & 0 & 0 & 0 & 1 & 0 \\
0 & 0 & -1 & 0 & 0 & 0 \\
0 & 0 & 0 & -1 & 0 & 0 \\
0 & 1 & 0 & 0 & 0 & 0 \\
1 & 0 & 0 & 0 & 0 & 0 
\end{array} \right), \quad \widetilde{\I}(g_3) = \left( \begin{array}{cccccc}
0 & 0 & 0 & 0 & 0 & 1 \\
0 & 0 & 0 & 1 & 0 & 0 \\
0 & -1 & 0 & 0 & 0 & 0\\
0 & 0 & -1 & 0 & 0 & 0 \\
1 & 0 & 0 & 0 & 0 & 0 \\
0 & 0 & 0 & 0 & 1 & 0 
\end{array} \right).
\end{equation}}}
There are three non-equivalent Bravais lattices in $\R^6$ left invariant by $\I$, classified in \cite{levitov}, namely the simple cubic (SC), body-centered cubic (BCC) and face-centered cubic lattice (FCC):
\begin{equation*}
\La_{SC} = \left\{ \bfx =(x_1, \ldots, x_6) : x_i \in \Z \right\},
\end{equation*}
 \begin{equation*}
\La_{BCC} = \left\{ \bfx = \frac{1}{2}(x_1, \ldots, x_6) : x_i \in \Z, \; x_i = x_j \; mod \; 2, \forall i,j=1,\ldots,6 \right\},
\end{equation*}
\begin{equation*}
\La_{FCC} = \left\{ \bfx = \frac{1}{2}(x_1, \ldots, x_6) : x_i \in \Z, \; \sum_{i=1}^6 x_i = 0 \; mod \; 2 \right\}.
\end{equation*}
The point group of these three lattices coincide, and it is referred to as the \emph{hyperoctahedral group} in six dimensions, and denoted by $B_6$. $\It$ leaves invariant two three-dimensional spaces, usually denoted as $E^{\parallel}$ (the physical space) and $E^{\perp}$ (the orthogonal space), both totally irrational with respect to any of the hypercubic lattices in $\R^6$. The matrix $R \in O(6)$ given by 
{\small{
\begin{equation}\label{projection_matrix}
R =  \frac{1}{\sqrt{2(2+\tau)}} \left( \begin{array}{cccccc}
\tau & 1 & 0 & \tau & 0 & 1 \\
0 & \tau & 1 & -1 & \tau & 0 \\
-1 & 0 & \tau & 0 & -1& \tau \\
0 & -\tau & 1 & 1 & \tau & 0 \\
\tau & -1 & 0 & -\tau & 0 & 1 \\
1 & 0 & \tau & 0 & -1 & -\tau
\end{array} \right) .
\end{equation}}}
decomposes $\It$ into irreducible representations, i.e. 
\begin{equation}\label{R}
\hat{\I}:= R^{-1}\It R = \rho_3 \oplus \rho_3',
\end{equation}
with $\rho_3 \simeq T_1$ and $\rho_3' \simeq T_2$. The explicit forms of $\rho_3$ and $\rho_3'$ are given in Table \ref{irreps_ico}. If $\pi^{\parallel} : \R^6 \rightarrow E^{\parallel}$ denotes the projection operator into the parallel space (and $\pi^{\perp}$ the corresponding orthogonal projection), we have, by linear algebra:
\begin{equation}\label{inv}
R^{-1} = \left( \begin{array}{c}
\pi^{\parallel} \\
\pi^{\perp} 
\end{array} \right),
\end{equation}
This setup allows the construction of icosahedral quasicrystals via the cut-and-project scheme:
\begin{equation*}
\begin{aligned}
E^{\parallel}\;  \stackrel{\pi^{\parallel}}{\longleftarrow} \; E^{\parallel} & \oplus E^{\perp} \; \stackrel{\pi^{\perp}}{\longrightarrow}\; E^{\perp} \\  
& \cup \\
& \La 
\end{aligned}
\end{equation*}
where $\La$ is one of the hypercubic lattices in $\R^6$. Specifically, we follow \cite{senechal} and consider as window $W$ the projection into the orthogonal space of the Voronoi cell $\mathcal{V}(\mathbf{0})$ of the origin, i.e. $W = \pi^{\perp}(\mathcal{V}(\mathbf{0}))$. Then the model set (cf. \cite{moody})
\begin{equation}\label{model_set_ico}
\Sigma(W) := \left\{ \pi^{\parallel}(\bfv) : \bfv \in \La, \; \pi^{\perp}(\bfv) \in W \right\}
\end{equation} 
defines a quasicrystal in $\R^3$ with icosahedral symmetry \cite{senechal, zappa1, paolo}; in a similar way,  icosahedral tilings of the space can be obtained with the dualisation method \cite{senechal, david, paolo}. 

In order to define structural transitions between icosahedral model sets with the Schur rotation method, we consider the classification of the crystallographic representations of $\I$ embedded in $B_6$ and the subsequent subgroup structure analysis provided by \cite{zappa1}. Specifically, the crystallographic representations of $\I$ form a unique conjugacy class of subgroups of $B_6$, which we denote by $\C_{B_6}(\It)$, whose order is 192, and whose representative is given by \eqref{gen}. Let $\G$ be a maximal subgroup of the icosahedral group $\I$, namely the tetrahedral group $\T$ or the dihedral groups $\D_{10}$ and $\D_6$, and let $\widetilde{\G}$ be a crystallographic representation of $\G$ embedded into the hyperoctahedral group $B_6$. Without loss of generality, we consider $\widetilde{\G}$ as a subgroup of the crystallographic representation $\widetilde{\I}$ given in \eqref{gen}; let $\C_{B_6}(\Gt)$ denote the conjugacy class of $\Gt$ in $B_6$. We have the following (for the proof, see \cite{zappa1}): 

\begin{prop}\label{reps}
 Let $\G$ be a maximal subgroup of $\I$. Then for every $P \in \C_{B_6}(\Gt)$ there exist \emph{exactly} two crystallographic representations of $\I$, $\h_1, \h_2 \in \C_{B_6}(\widetilde{\I})$,  such that $P = \h_1 
 \cap \h_2$.
\end{prop} 

We point out that in the case of achiral icosahedral symmetry, the symmetry group to be considered is the Coxeter group $H_3$ \cite{coxeter}. Due to the isomorphism $H_3 \simeq \I \times \Z_2$, the representation theory of $H_3$ easily follows from this direct product structure. In particular, the representation $\h_3 := \It \otimes \Gamma$, where $\Gamma = \{1, -1\}$ is the non-trivial irreducible representation of $\Z_2$, is a six-dimensional crystallographic representation of $H_3$ in the sense of Definition \ref{cryst_def}. Therefore, all the analysis developed in the following still holds for achiral model sets and arrays.      

\begin{table}[!t]
\begin{center}
\begin{tabular}{|c c c|}
\hline
Generator & Irrep $\rho_3 \simeq T_1$ & Irrep $\rho_3' \simeq T_2$ \\
\hline
$g_2$ & $\frac{1}{2} \left( \begin{array}{ccc}
\tau-1 & 1 & \tau \\
1 & -\tau & \tau-1 \\
\tau & \tau-1 & -1
\end{array} \right)$ &  $\frac{1}{2} \left( \begin{array}{ccc}
\tau-1 & -\tau & -1 \\
-\tau & -1 & \tau-1 \\
-1 & \tau-1 & -\tau 
\end{array} \right)$ \\
\hline
$g_3$  & $\frac{1}{2} \left( \begin{array}{ccc}
\tau & \tau-1 & 1 \\
1-\tau & -1 & \tau \\
1 & -\tau & 1-\tau
\end{array} \right)$ &  $\frac{1}{2} \left( \begin{array}{ccc}
-1 & 1-\tau & -\tau \\
\tau-1 & \tau & -1 \\
\tau & -1 & 1-\tau 
\end{array} \right)$ \\
\hline
\end{tabular}
\end{center}
\caption{\label{irreps_ico} Explicit forms of the irreps $\rho_3$ and $\rho_3'$ with $\widetilde{\I} \simeq \rho_3 \oplus \rho_3'$.}
\end{table}

\section{Schur rotations between icosahedral structures}\label{schur}

In this section we define structural transitions between icosahedral quasicrystals and finite point arrays using the Schur rotation method, and prove the connection with the Bain strain method described in \cite{paolo}.

Let $\Gt$ be a representation of a maximal subgroup $\G$ of $\I$, which is a matrix subgroup of $\It$ given in \eqref{gen}. By Proposition \ref{reps}, there exists a unique crystallographic representation of $\I$ in $B_6$, which we denote by $\widetilde{\I}_{\G}$, such that $\Gt$ is a subgroup of $\It$ and $\It_{\G}$, i.e. $\widetilde{\G} = \widetilde{\I} \cap \widetilde{\I}_{\G}$. The matrix $R$ in \eqref{projection_matrix}, which reduces $\It$ into irreps  as in \eqref{R}, decomposes the representation $\Gt$ as follows:
\begin{equation}\label{g_rep}
\hat{\G}:= R^{-1}\Gt R = \G_1 \oplus \G_2,
\end{equation}
where $\G_1$ and $\G_2$ are matrix subgroups of the irreps $\rho_3$ and $\rho_3'$ given in Table \ref{irreps_ico}, respectively. Notice that $\G_1$ and $\G_2$ are not necessarily irreducible representations of $\Gt$. 

The matrix $R$ in general does not reduce the representation $\It_{\G}$, since the subspaces $E^{\parallel}$ and $E^{\perp}$, which are invariant under $\It$, are not necessarily invariant under $\It_{\G}$. Let us denote by $R_{\G} \in O(6)$ the orthogonal matrix that reduces $\It_{\G}$ into irreps, i.e.
\begin{equation*}
\hat{\I}_{\G} := R_{\G}^{-1} \It_{\G} R_{\G} \simeq T_1 \oplus T_2,
\end{equation*}
where $T_1$ and $T_2$ are the two non-equivalent three-dimensional irreps of $\I$. This matrix carries the bases of a physical and a parallel space which are invariant under $\It_{\G}$. We denote these spaces by $E^{\parallel}_{\G}$ and $E^{\perp}_{\G}$, respectively, and we write $\pi^{\parallel}_{\G}$ and $\pi^{\perp}_{\G}$ for the corresponding projections. By \eqref{inv}, we have
\begin{equation*}
R_{\G}^{-1} = \left( \begin{array}{c}
\pi^{\parallel}_{\G} \\
\pi^{\perp}_{\G} 
\end{array} \right).
\end{equation*}

The matrix $R_{\G}$ is in general not unique. With a suitable choice of the basis vectors constituting the columns of $R_{\G}$, we assume $\det(R_{\G})$ and $\det(R)$ have the same sign, i.e. $R$ and $R_{\G}$ belong to the same connected component of $O(6)$.  Furthermore, since $\Gt$ is a common subgroup of $\It$ and $\It_{\G}$,  it is possible to choose $R_{\G} \in O(6)$ such that  $\hat{\G} = \hat{\I} \cap \hat{\h}_{\G}$, i.e.
\begin{equation}\label{rel_g}
R^{-1} G R = R_{\G}^{-1} G R_{\G}, \; \forall G \in \Gt \Rightarrow (R_{\G}R^{-1})^{-1} G (R_{\G}R^{-1}) = G, \; \forall G \in \Gt. 
\end{equation} 
Therefore $R_{\G}R^{-1}$ belongs to the centraliser of $\Gt$ in $GL(6,\R)$, i.e. the set
\begin{equation*}
\z(\Gt, \R) := \{ A \in GL(6,\R) : A G = G A, \; \forall G \in \Gt \}\,.
\end{equation*}
Hence there exists a matrix $M_{\G} \in \mathcal{Z}(\Gt,\R) \cap O(6)$, denoted as the \emph{Schur operator} related to $\G$, such that $R_{\G} = M_{\G} R$. Since $R$ and $R_{\G}$ have determinants with equal signs by assumption, we have that $\det(M_{\G}) > 0$, hence $M_{\G}$ is a rotation in  $SO(6)$. Let us consider a path
\begin{equation}\label{schur_rotation}
M_{\G}(t) : [0,1] \longrightarrow \mathcal{Z}(\Gt,\R) \cap SO(6)
\end{equation}
that connects $M_{\G}$ to the identity matrix $I_6$, i.e. $M_{\G}(0) = I_6$ and $M_{\G}(1) = M_{\G}$. Such a path is referred to as the \emph{Schur rotation} associated with $\Gt$. The name comes from Schur's Lemma in Representation Theory, that gives constraints on the matrices that commute with a representation of a group \cite{jones}. In Section  \ref{schur_comp} we prove the existence and determine the explicit forms of \eqref{schur_rotation} for all the maximal subgroups of the icosahedral group.

Let us consider the path $R_{\G}(t) : [0,1] \rightarrow O(6)$ defined by $R_{\G}(t) := M_{\G}(t)R$. For every $t \in [0,1]$ the matrix $R_{\G}(t)$ encodes the basis of a physical space $E^{\parallel}_t$ and an orthogonal space $E^{\perp}_t$ that carry the representations $\G_1$ and $\G_2$ of $\G$ as in \eqref{g_rep} since
\begin{equation}\label{comm_G}
R_{\G}(t)^{-1} \Gt R_{\G}(t) = R^{-1} M_{\G}(t)^{-1} \Gt M_{\G}(t) R = R^{-1} \Gt R = \G_1 \oplus \G_2. 
\end{equation}
In particular, we have $E^{\parallel}_t = M_{\G}(t) E^{\parallel}$ and $E^{\parallel}_0 \equiv E^{\parallel}$, $E^{\parallel}_1 \equiv E^{\parallel}_{\G}$  (and similarly for the orthogonal spaces). For $t \in [0,1]$, the projections $\pi^{\parallel}_t : \R^6 \rightarrow E^{\parallel}_t$ and $\pi^{\perp}_t : \R^6 \rightarrow E^{\perp}_t$ are given by (compare with \eqref{inv}):
\begin{equation}\label{pt}
 \left( \begin{array}{c}
\pi^{\parallel}_{t} \\
\pi^{\perp}_{t} 
\end{array} \right) = R_{\G}^{-1}(t) = R^{-1}M_{\G}(t)^{-1} = 
 \left( \begin{array}{c}
\pi^{\parallel} \\
\pi^{\perp} 
\end{array} \right)M_{\G}(t)^{-1} =  \left( \begin{array}{c}
\pi^{\parallel} M_{\G}(t)^{-1}\\
\pi^{\perp} M_{\G}(t)^{-1} 
\end{array} \right).
\end{equation}

\begin{figure}[!t]
\centering
\begin{tikzpicture}[scale = 0.8]

\coordinate (Origin)   at (0,0);
    \coordinate (XAxisMin) at (-5,0);
    \coordinate (XAxisMax) at (8,0);
    \coordinate (YAxisMin) at (0,-2);
    \coordinate (YAxisMax) at (0,5);
    \draw [thin, gray,-latex] (XAxisMin) -- (XAxisMax);
    \draw [thin, gray,-latex] (YAxisMin) -- (YAxisMax);

\foreach \x in {-5,-4,...,8}{
      \foreach \y in {-2,-1,...,5}{
        \node[draw,circle,inner sep=1pt,fill] at (\x,\y) {};
}
}

\draw [blue] (-4,-2.472) -- (8.4,5.191);
\draw [red] (1.5, -2.427) -- (-3.5,5.663);

\draw [dashed, blue] (-2.2,-3.028) -- (4.2, 5.78);
\draw [dashed, red] (3.2,-2.325) -- (-5.2, 3.77);

\draw [blue] (2, 1.236) to[bend right] (1.2, 1.6516);

\node at (1.5, 1.3) {$\beta$};

\end{tikzpicture}
\caption[Illustration of the Schur rotation for a one-dimensional quasicrystal.]{Illustration of the Schur rotation for a one-dimensional quasicrystal. The physical space (straight line in red) and the orthogonal space (straight line in blue) undergo a rotation of an angle $\beta$, resulting in the new physical and orthogonal spaces (dashed lines). The two-dimensional lattice remains fixed throughout the rotation.}
\label{schur_fig} 
\end{figure}
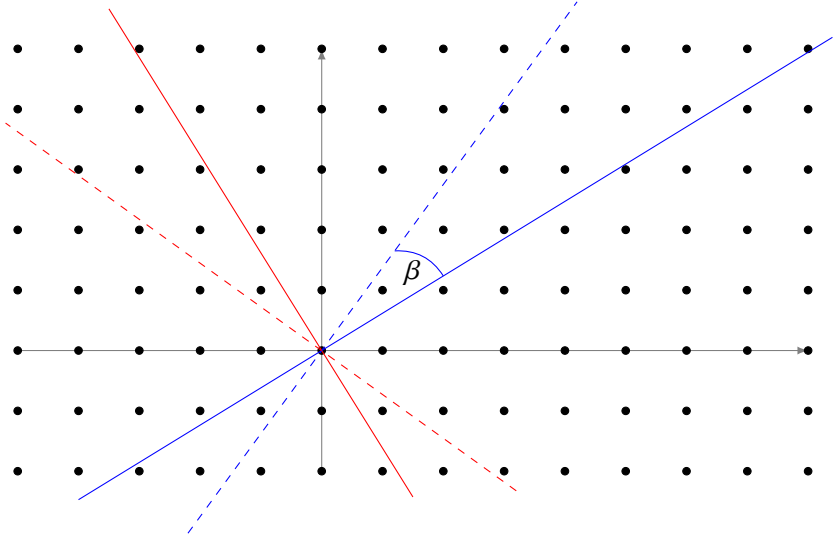

With this setup, we can define structural transitions between icosahedral quasicrystals that keep the symmetry encoded by $\G$ preserved. Specifically, let $\La$ be one of the three hypercubic lattices in $\R^6$ described in Section \ref{embedding}, and let $\Sigma(W)$ be the icosahedral model set as in \eqref{model_set_ico}. Let us then consider, for all $t \in [0,1]$, the projection $ W_t := \pi^{\parallel}_t((\mathcal{V}(\mathbf{0}))$ of $\mathcal{V}(\mathbf{0})$ into the space $E^{\perp}_t$. We define the family of model sets
\begin{equation}\label{sigma_t}
\Sigma_t \equiv \Sigma(W_t) := \left\{ \pi^{\parallel}_t (\bfv) : \bfv \in \La, \; \pi^{\perp}_t(\bfv) \in W_t \right\}. 
\end{equation}
By construction, $\Sigma_0 \equiv \Sigma(W)$ and $\Sigma_1$ possess icosahedral symmetry, whereas the intermediate states $\Sigma_t$, for $t \in (0,1)$, display $\G$-symmetry since, by \eqref{comm_G}:
\begin{equation}\label{comm_G_t}
\pi^{\parallel}_t(\Gt \bfv) = \G_1 \pi^{\parallel}_t (\bfv), \quad \forall t \in (0,1).
\end{equation}
Hence, the Schur rotation $M_{\G}(t)$ as in \eqref{schur_rotation} defines a continuous transformation of the model set $\Sigma(W)$ into another icosahedral quasilattice, where $\G$-symmetry is preserved. We point out that, in the higher dimensional space, the lattice $\La$ is \emph{fixed} and the transformation is induced by the rotation of the physical and orthogonal spaces (see Figure \ref{schur_fig}). The angle(s) of rotation correspond(s) to the degree(s) of freedom of the transformation, and can be chosen as the \emph{order parameter(s)} of the transition in the framework of the Landau theory \cite{sethna}. 

\paragraph{Transitions of finite icosahedral point sets.} In the context of virology and carbon chemistry, the arrangements of viral proteins and carbon atoms in fullerenes are modeled via finite point sets (arrays) with icosahedral symmetry. The method developed here can also be applied to analyse structural transitions between icosahedral arrays, creating finite point sets via projection, as opposed to the infinite ones generated by the cut-and-project scheme, at every time $t$ of the transformation. Indeed, let $\C = \{ \pi^{\parallel}(\bfv_i) : \bfv_i \in \La, i = 1,\ldots, n \}$ be a finite point set in $E^{\parallel}$, obtained via the projection of points of a hypercubic lattice $\La$ in $\R^6$. Let us assume that $\C$ is closed under the action of the irrep $\rho_3$ of $\I$ (cf.~\eqref{R}), i.e. $\rho_3 \C \subseteq \C$. The projection operators $\pi^{\parallel}_t$ given in \eqref{pt} can be used to define a family of arrays $\C_t$, for $t \in [0,1]$, given by:
\begin{equation}\label{clusters}
\C_t := \left\{ \pi^{\parallel}_t(\bfv_i) : \bfv_i \in \C, i =1, \ldots, n \right\}.
\end{equation}
It follows from \eqref{comm_G_t} that the point sets $\C_t$ are invariant under the representation $\G_1$ of $\G$ (cf.~\eqref{g_rep}) for all $t \in (0,1)$, and moreover possess icosahedral symmetry for $t=0$ and $t=1$. We refer to \cite{zappa2} for a detailed method for the construction of finite nested icosahedral point sets via projection, in connection with the structure of viral capsids. 

\paragraph{Connection with the Bain strain method.} In crystallography and condensed matter physics, the concept of Bain strain relates to deformations of three-dimensional lattices that keep some symmetry preserved, described by a common subgroup of the point groups of the lattices which constitute the initial and final states \cite{zanzotto}. Indelicato \emph{et al.} \cite{paolo, giuliana} provided a higher-dimensional generalisation of the Bain strain for lattices in $\R^n$. In this context, given two lattices $\La_0$ and $\La_1$  with generator matrices $B_0$ and $B_1$, respectively, and a subgroup $\h$ of $\Po(\La_0)$ and $\Po(\La_1)$, a transition between $\La_0$ and $\La_1$ with symmetry $\h$ is a path $B(t) : [0,1] \rightarrow GL(n,\R)$ such that, if $\La_t$ denotes the intermediate lattice with generator matrix $B(t)$, then $\h \subseteq \Po(\La_t)$, for all $t \in [0,1]$. If $\La_0$ and $\La_1$ are 6D hypercubic lattices, and $\h = \Gt$ a maximal subgroup of $\It$, then $B(t)$ induces a continuous transformation 
\begin{equation}\label{sigma_t_2}
\widetilde{\Sigma}_t := \left\{ \pi^{\parallel}(\bfv_t) : \pi^{\perp}(\bfv_t) \in \pi^{\perp}(\mathcal{V}_t(\mathbf{0})) \right\}, 
\end{equation}   
where $\bfv_t := B(t)\boldsymbol{m}$, $\boldsymbol{m} \in \Z^6$, is a point in the intermediate lattice $\La_t$, and $\mathcal{V}_t(\mathbf{0})$ denotes the Voronoi cell of $\La_t$ at the origin. The symmetry identified by $\G$ is preserved since $\G_1 \pi^{\parallel}(\bfv_t) = \pi^{\parallel}(\Gt \bfv_t)$, and $\Gt \bfv_t \in \La_t$ since $\Gt \subseteq \Po(\La_t)$ for all $t \in [0,1]$. We notice that in this approach the lattice undergoes a transformation, whereas the physical and orthogonal spaces remain fixed. 

As already pointed out in \cite{paolo}, the Schur rotation and the generalised Bain strain are related. This can easily be proved with the mathematics developed so far. In particular, if $M_{\G}(t)$ is a Schur rotation associated with $\G$ as in \eqref{schur_rotation}, let us define the path $\hat{B}(t) : [0,1] \rightarrow GL(6,\R)$ as  
\begin{equation}
\hat{B}(t):= M_{\G}(t)^{-1} B_0.
\end{equation}
We have, using \eqref{pt},
\begin{equation*}
\pi^{\parallel}\left(\hat{B}(t)\right) = \pi^{\parallel}\left(M_{\G}(t)^{-1} B_0\right) = \pi^{\parallel}_t(B_0),
\end{equation*} 
and similarly for $\pi^{\perp}_t$. Therefore  
\begin{equation*}
\widetilde{\Sigma}_t = \left\{ \pi^{\parallel}(\bfv_t) : \pi^{\perp}(\bfv_t) \in \pi^{\perp}(\mathcal{V}_t(\mathbf{0})) \right\} = \left\{ \pi^{\parallel}_t(B_0\boldsymbol{m}) : \pi^{\perp}_t(B_0\boldsymbol{m}) \in \pi^{\perp}_t(\mathcal{V}(\mathbf{0})) \right\} = \Sigma_t,
\end{equation*}
and moreover $\Gt \subseteq \Po(\La_t)$, since $\Po\left(\hat{B}(t)\right) = M_{\G}^{-1}(t) \Po(B_0) M_{\G}(t)$ (this is true since, for every lattice $\La$ in $\R^n$ with generator matrix $B$, $\Po(RB) = R\Po(B)R^{-1}$, for $R \in O(n)$ \cite{zanzotto})  and $M_{\G}(t) \in \Z(\Gt,\R)$ for all $t$. Hence the Schur rotation is equivalent to a Bain strain transformation between \emph{congruent} lattices, i.e. whose bases are related via a rotation \cite{conway}. The advantage of the former is that the use of Schur's Lemma and tools from representation theory can be used in the computation and allow a characterisation of such transitions in a purely group theoretical framework.

\section{Computations and applications}\label{schur_comp}

In this section we compute the Schur rotations for the maximal subgroups of the icosahedral group, and discuss applications and specific examples. It follows from Section \ref{schur} that the crucial point is the computation of the matrix groups $\z(\Gt,\R) \cap SO(6)$, where $\Gt \subseteq \It$ is a representation of $\G$ in $B_6$. To this aim, we first focus on the group $\z(\hat{\G},\R) \cap SO(6)$, which consists of all the rotations in $SO(6)$ that commute with the matrices constituting the \emph{reduced} representation $\hat{\G}$. A matrix in this group can be easily computed using Schur's Lemma \cite{jones}; the group  $\z(\Gt,\R) \cap SO(6)$ then easily follows since the following holds (see \cite{artin}):
\begin{equation}\label{t_1}
\z(\hat{\G}, \R) = \z(R^{-1} \Gt R, \R) = R^{-1} \z(\Gt,\R) R.
\end{equation}
We now consider in detail the computations and examples for each maximal subgroup of the icosahedral group.
 
\subsection{Tetrahedral group $\T$}

The tetrahedral group $\T$ is the rotational symmetry group of a tetrahedron, generated by a two-fold rotation $g_2$ and a three-fold rotation $g_{3d}$ such that $g_2^2=g^{3}_{3d} = (g_2g_{3d})^3 = e$. It is isomorphic to the alternating group $\mathfrak{A}_4$ and its character table is given by (cf. \cite{jones}):
\begin{center}
{\small{
\begin{tabular}{l|c c c c }
Irrep & $\mathcal C(e)$ & $4C_3$ & $4C_3^2$ & $3C_2$ \\
\hline
A & 1 & 1 & 1 & 1 \\
$E_1$ & 1 & $\omega$ & $\omega^2$ & 1 \\
$E_2$ & 1 & $\omega^2$ & $\omega$ & 1 \\
T & 3 & 0 & 0 & -1
\end{tabular}}}
\end{center}
where $ \omega = e^{\frac{2\pi i}{3}}$. Note that the representations $E_1$ and $E_2$ are complex, while their direct sum $E := E_1 \oplus E_2$  is real and  irreducible in $GL(2,\R)$. An explicit representation $\Tt$ of $\T$, which is a subgroup of $\It$, is given by
{\small{
\begin{equation*}
\Tt = \left< \left( \begin{array}{cccccc}
0 & 0 & 0 & 0 & 0 & 1 \\
0 & 0 & 0 & 0 & 1 & 0 \\
0 & 0 & -1 & 0 & 0 & 0 \\
0 & 0 & 0 & -1 & 0 & 0 \\
0 & 1 & 0 & 0 & 0 & 0 \\
1 & 0 & 0 & 0 & 0 & 0 
\end{array} \right), \left( \begin{array}{cccccc}
0 & 1 & 0 & 0 & 0 & 0 \\
0 & 0 & 0 & -1 & 0 & 0 \\
0 & 0 & 0 & 0 & 0 & -1\\
-1 & 0 & 0 & 0 & 0 & 0 \\
0 & 0 & 1 & 0 & 0 & 0 \\
0 & 0 & 0 & 0 & -1 & 0 
\end{array} \right) \right>\,.
\end{equation*}}}
This representation can be found with the aid of the software \texttt{GAP} \cite{gap}, which is designed for problems in computational group theory (see \cite{zappa1} for more details). All the subsequent group theoretical computations were performed in \texttt{GAP}. 

The matrix $R$ as in \eqref{projection_matrix} is such that
\begin{equation}\label{tet_irreps}
\hat{\T} :=R^{-1} \Tt R = \Gamma_1 \oplus \Gamma_2,
\end{equation}
where $\Gamma_1$ and $\Gamma_2$ are matrix subgroups of $\rho_3$ and $\rho_3'$ as in \eqref{R}, respectively,  and both are equivalent to the irrep $T$ of $\T$. Due to this equivalence, there exists a matrix $Q \in GL(3,\R)$ such that $Q^{-1} \Gamma_2 Q = \Gamma_1$.  The explicit forms of $\Gamma_1$, $\Gamma_2$ and $Q$ are given in Table \ref{tet_reps_table}. Note that $Q$ can be chosen to be orthogonal. Let us define $\hat{Q}:=I_3 \oplus Q \in O(6, \R)$, where $I_3$ denotes the $3 \times 3$ identity matrix; then we have
\begin{equation}\label{t_2}
\overline{\T}:= \hat{Q}^{-1} \hat{\T} \hat{Q} = \Gamma_1 \oplus \Gamma_1.
\end{equation}
We consider the set $\z(\overline{\T},\R) \cap SO(6)$. Writing a matrix $N$ in this group as 
\begin{equation*}
N = \left( \begin{array}{cc}
N_1 & N_2 \\
N_3 & N_4 
\end{array} \right),
\end{equation*}
where $N_i$ are $3 \times 3$ matrices, for $i =1, \ldots, 4$, we impose $N \overline{\T} = \overline{\T} N$, i.e. $N (\Gamma_1 \oplus \Gamma_1) = (\Gamma_1 \oplus \Gamma_1) N$.
Using Schur's Lemma and imposing orthogonality, we obtain 
\begin{equation*}
N = N(\beta) = \left( \begin{array}{cc}
\cos(\beta)I_3 & -\sin(\beta)I_3 \\
\sin(\beta)I_3 & \cos(\beta)I_3
\end{array} \right),
\end{equation*}
where $\beta$ belongs to the unit circle $S^1$. Notice that $N(\alpha)N(\beta) = N(\alpha+\beta)$. Putting together \eqref{t_2} and \eqref{t_1} we obtain
\begin{equation*}
 \z(\Tt,\R) \cap SO(6) = \left\{ (R\hat{Q}) N(\beta) (R\hat{Q})^{-1} :   N(\beta) = \left( \begin{array}{cc}
\cos(\beta)I_3 & -\sin(\beta)I_3 \\
\sin(\beta)I_3 & \cos(\beta)I_3
\end{array} \right), \; \beta \in S^1 \right\}.
\end{equation*}
It follows that the group $\z(\Tt,\R) \cap SO(6)$ is isomorphic to $S^1$, hence it is a compact and connected Lie group. Therefore, the angle $\beta \in S^1$ can be chosen as an order parameter for the transitions with tetrahedral symmetry. 

In order to compute the Schur rotations between icosahedral quasicrystals with $\T$-symmetry,  we need to fix the boundary conditions, i.e. imposing the end and the start of the transition to have icosahedral symmetry. In particular, we consider the crystallographic representation $\It_{\T}$ of $\I$ with the property that $\Tt = \It \cap \It_{\T}$:
{\small{
\begin{equation*}
\It_{\T} = \left< \left( \begin{array}{cccccc}
0 & 0 & 0 & -1 & 0 & 0 \\
0 & 0 & 0 & 0 & 0 & -1 \\ 
0 & 0 & -1 & 0 & 0 & 0 \\ 
-1 & 0 & 0 & 0 & 0 & 0 \\ 
0 & 0 & 0 & 0 & -1 & 0 \\ 
0 & -1 & 0 & 0 & 0 & 0
\end{array} \right),  \left( \begin{array}{cccccc}
0 & 0 & 0 & -1 & 0 & 0 \\
0 & 0 & 0 & 0 & -1 & 0 \\ 
0 & -1 & 0 & 0 & 0 & 0 \\ 
0 & 0 & 0 & 0 & 0 & 1 \\
0 & 0 & 1 & 0 & 0 & 0 \\ 
-1 & 0 & 0 & 0 & 0 & 0
\end{array} \right) \right>.
\end{equation*}}}
Given $M_{\T}(\beta) \in \z(\Tt,\R) \cap SO(6)$, we consider the matrix $R_{\T}(\beta) := M_{\T}(\beta) R \in O(6)$ and impose 
\begin{equation}\label{eq_dec}
R_{\T}(\beta)^{-1} \It_{\T} R_{\T}(\beta) \simeq T_1\oplus T_2 \,.
\end{equation}
We solve (\ref{eq_dec}) with respect to $\beta$; in other words, we look for angles $\hat{\beta} \in S^1$ such that the corresponding matrix $R_{\T}(\hat{\beta})$ decomposes into irreps of the representation $\It_{\T}$. Specifically, let $M_2$ and $M_3$ denote the generators of $\It_{\T}$, and let us define the matrices $K_j(\beta) := R_{\T}(\beta)^{-1}M_jR_{\T}(\beta)$, for $j = 2,3$. Condition \eqref{eq_dec} is then equivalent to the following system of $36$ equations: 
\begin{equation}\label{36}
\left\{ \begin{aligned}
& \left(K_2(\beta)\right)_{ij} = 0, \; \left(K_2(\beta)\right)_{ji} = 0 \\  
& \left(K_3(\beta)\right)_{ij} = 0, \;  \left(K_3(\beta)\right)_{ji} = 0 \; 
\end{aligned} \right.
\end{equation}
for $i = 1, 2,3$ and $j = 4,5,6$. The solutions of \eqref{36} are given by:
\begin{equation*}
\hat{\beta} \in \left\{ -\arctan\left(\frac{1}{2}\right), \; -\arctan\left(\frac{1}{2}\right)+\pi, \; \arctan(2), \; \arctan(2)-\pi \right\} =: S_{\T}. 
\end{equation*}
Hence the number of Schur operators associated with $\T$ is finite; the elements in $S_{\T}$ provide all the possible boundary conditions for the analysis of transitions with $\T$-symmetry between icosahedral order. Specifically, since $S^1$ is connected, we can consider any path $\beta(t) : [0,1] \rightarrow S^1$ that connects $0$ with $\hat{\beta} \in S_{\T}$, i.e. $\beta(0) = 0$ and $\beta(1) = \hat{\beta}$. Then the corresponding Schur rotation $M_{\T}(t)$ is given by $M_{\T}(\beta) \circ \beta(t) = M_{\T}(\beta(t)) : [0,1] \rightarrow \z(\Tt, \R) \cap SO(6)$. 
 
\begin{table}[!t]
\begin{center}
\begin{tabular}{|c c c|}
\hline
Generator & Irrep $\Gamma_1$ & Irrep $\Gamma_2$ \\
\hline
$g_2$ & $\frac{1}{2} \left( \begin{array}{ccc}
\tau-1 & 1 & \tau \\
1 & -\tau & \tau-1 \\
\tau & \tau-1 & -1
\end{array} \right)$ &  $\frac{1}{2} \left( \begin{array}{ccc}
\tau-1 & -\tau & -1 \\
-\tau & -1 & \tau-1 \\
-1 & \tau-1 & -\tau 
\end{array} \right)$ \\
$g_{3d}$  & $\frac{1}{2} \left( \begin{array}{ccc}
1-\tau & 1 & \tau \\
1 & \tau & 1-\tau \\
-\tau & \tau-1 & -1
\end{array} \right)$ &  $\frac{1}{2} \left( \begin{array}{ccc}
1-\tau & \tau & -1 \\
-\tau & -1 & 1-\tau \\
-1 & \tau-1 & \tau 
\end{array} \right)$ \\
\hline
& $Q  = \frac{1}{4} \left( \begin{array}{ccc}
3-\tau & 1 & \tau+2 \\
-\tau-2 & 3-\tau & 1 \\
-1 & -\tau-2 & 3-\tau
\end{array} \right)$ & \\
\hline
\end{tabular}
\end{center}
\caption{\label{tet_reps_table} Explicit forms of the irreps $\Gamma_1$ and $\Gamma_2$ of the tetrahedral group and the matrix $Q \in O(3)$ such that $Q^{-1} \Gamma_2 Q = \Gamma_1$ (cf. \eqref{tet_irreps}).}
\end{table}

\begin{figure}[!t]
\includegraphics[scale = 0.22]{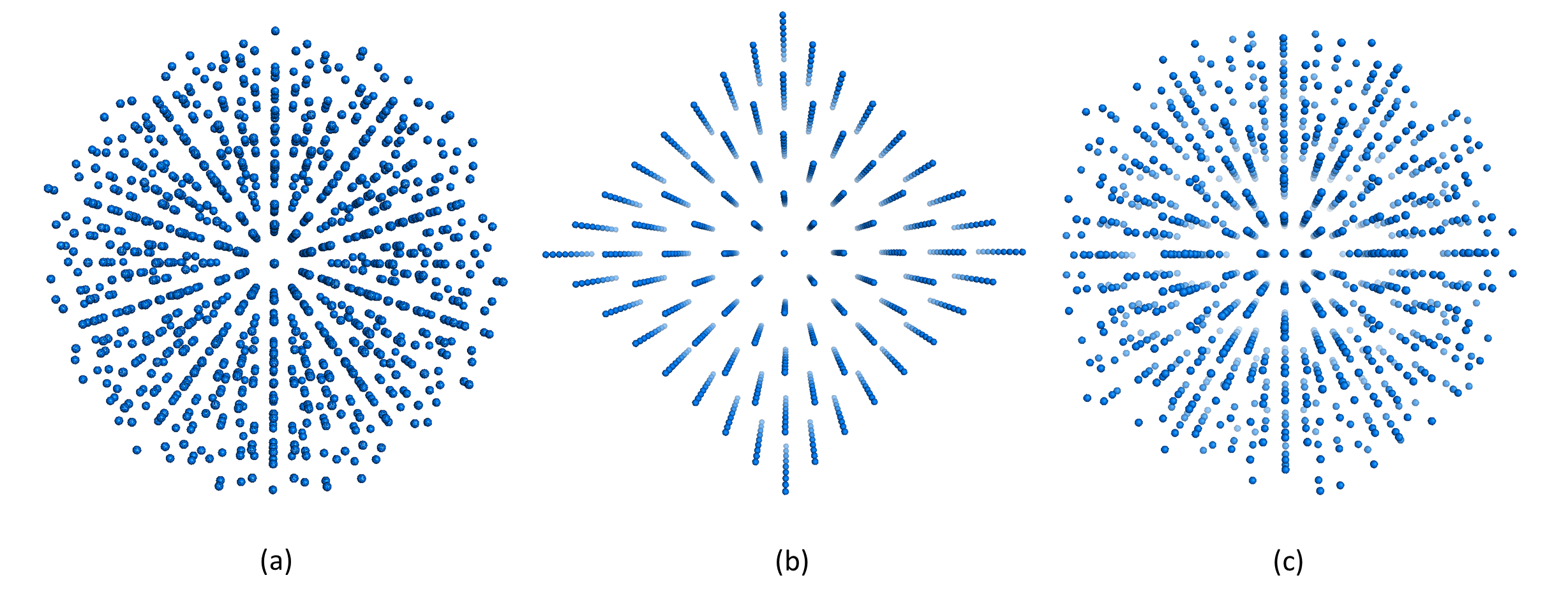}
\caption[Example of a transition with tetrahedral symmetry.]{Example of a transition with tetrahedral symmetry. The model sets in (a) and (c) correspond to the starting and final states, respectively, and display icosahedral symmetry. The intermediate state in (b) exhibits octahedral symmetry and defines a three-dimensional cubic lattice. }
\label{trans_tet}
\end{figure}

\paragraph{Example: tetrahedral transition with an intermediate cubic lattice.} We consider as an explicit example of a tetrahedral transition the path $\beta(t) = \hat{\beta}t$, that connects $0$ with $\hat{\beta} = -\arctan\left(\frac{1}{2}\right)$. The lattice $\La$ in $\R^6$ is taken as the simple cubic lattice with the standard basis in $\R^6$. The matrix $R_{\T}(t) = M_{\T}(\hat{\beta}t) R$ encodes the projections $\pi^{\parallel}_t$ and $\pi^{\perp}_t$ as in \eqref{pt}, that define the   family of model sets $\Sigma_t$ as in \eqref{sigma_t}. In Figure \ref{trans_tet} we show a patch of the resulting quasilattices for $t = 0$, $0.5$ and $1$. These are very interesting results; indeed,  the starting and final states display icosahedral aperiodicity, as expected by the boundary conditions, while for $t = 0.5$ the corresponding structure is actually a three-dimensional lattice, i.e. it is periodic. Hence such a transition has an intermediate periodic order, which is in accordance to the previous result by Kramer \cite{kramer}. From a group theoretical point of view, this implies that there exists a subgroup of $B_6$ isomorphic to the octahedral group $\mathcal{O}$ (i.e. the symmetry group of a cube, with order $48$), which contains $\Tt$ as a subgroup.   

\subsection{Dihedral group $\D_{10}$}\label{d10}

\begin{figure}[!t]
\centering
\includegraphics[scale = 0.13]{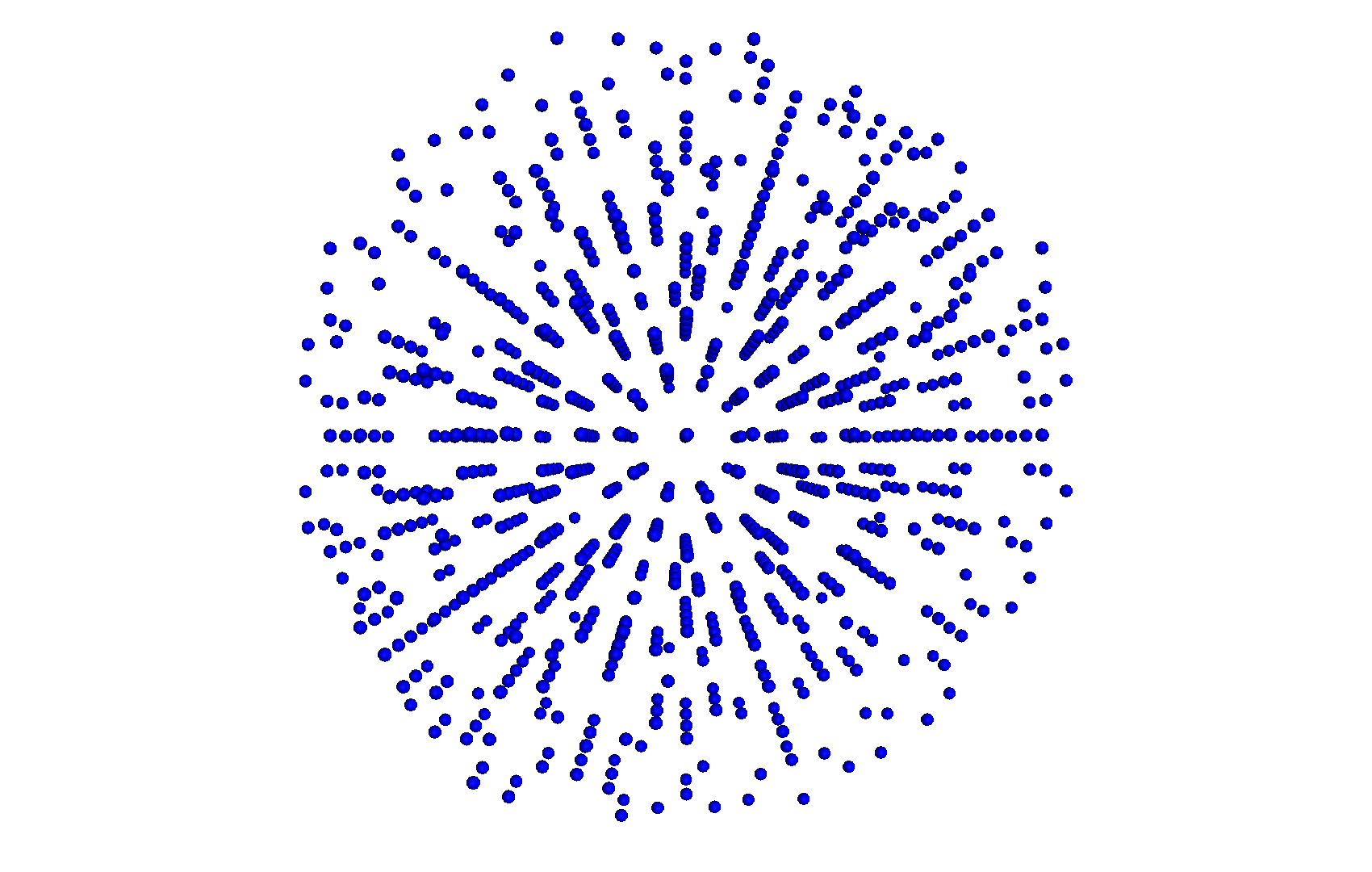}
\caption[Patch of a quasilattice with $\D_{10}$-symmetry.]{Patch of a quasilattice with $\D_{10}$-symmetry, obtained from the Schur rotation associated with $\D_{10}$ and corresponding to the intermediate state $\beta = \frac{\pi}{4}$. }
\label{D10_lattice}
\end{figure}

The dihedral group $\D_{10}$ is generated by two elements $g_{2d}$ and $g_{5d}$ such that $g_{2d}^2 = g_{5d}^5 = (g_{2d}g_{5d})^2 = e$. Its character table is as follows \cite{jones}:
\begin{center}
{\small{
\begin{tabular}{l|c c c c }
Irrep & $E$ & $2C_5$ & $2C_5^2$ & $5C_2$ \\
\hline
$A_1$ & 1 & 1 & 1 & 1 \\
$A_2$ & 1 & 1 & 1 & -1 \\
$E_1$ & 2 & $\tau-1$ & $-\tau$ & 0 \\
$E_2$ & 2 & $-\tau$ & $\tau-1$ & 0
\end{tabular}}}
\end{center} 
An explicit representation $\Dt_{10}$ as a matrix subgroup of $\It$ is given by 
{\small{
\begin{equation*}
\Dt_{10} = \left< \left( \begin{array}{cccccc}
0 & 0 & 0 & 0 & 0 & -1 \\
0 & -1 & 0 & 0 & 0 & 0 \\
0 & 0 & 0 & 1 & 0 & 0 \\
0 & 0 & 1 & 0 & 0 & 0 \\
0 & 0 & 0 & 0 & -1 & 0 \\
-1 & 0 & 0 & 0 & 0 & 0 
\end{array} \right), \left( \begin{array}{cccccc}
0 & 0 & 0 & 0 & 0 & 1 \\
0 & 1 & 0 & 0 & 0 & 0 \\
0 & 0 & 0 & 0 & -1 & 0\\
-1 & 0 & 0 & 0 & 0 & 0 \\
0 & 0 & 0 & 1 & 0 & 0 \\
0 & 0 & 1 & 0 & 0 & 0 
\end{array} \right) \right>.
\end{equation*}}}
In order to compute the Schur rotations associated with $\D_{10}$, we proceed in a similar way as in the tetrahedral case. The projection matrix $R$ in \eqref{projection_matrix} decomposes $\Dt_{10}$ as
\begin{equation}\label{d10_rel}
\hat{\D}_{10}:= R^{-1} \Dt_{10} R = D_1 \oplus D_2,
\end{equation}
where $D_1$ and $D_2$ are matrix subgroups of $\rho_3$ and $\rho_3'$, respectively, that are \emph{reducible} representations of $\D_{10}$. In particular, from its character table we have that $D_1 \simeq A_2 \oplus E_1$ and $D_2 \simeq A_2 \oplus E_2$ in $GL(3, \R)$. In order to find the Schur operators for $\D_{10}$, we first reduce $D_1$ and $D_2$ into irreps, using tools from the representation theory of finite groups (see, for example, \cite{fulton}). In particular, we determine two orthogonal matrices, $P_1$ and $P_2$, such that
\begin{equation}\label{d10_reps}
\hat{D}_1 := P_1^{-1} D_1 P_1 \simeq A_2 \oplus E_1, \quad \hat{D}_2 := P_2^{-1}D_2 P_2 \simeq A_2 \oplus E_2.
\end{equation}
The explicit forms of $D_1$, $D_2$, $P_1$ and $P_2$ are given in Table \ref{d10_reps}. The matrix $Z := P_1 \oplus P_2$ is such that (cf. \eqref{d10_rel}):
\begin{equation*}
Z^{-1}(R^{-1}\Dt_{10}R)Z =  Z^{-1} \hat{\D}_{10} Z = \hat{D}_1 \oplus \hat{D}_2 =: \overline{\D}_{10}. 
\end{equation*}
By Schur's Lemma, a matrix $M \in \z(\overline{\D}_{10}, \R) \cap SO(6)$ must be of the form
{\small{
\begin{equation*}
M = M(\beta)  =  \left( \begin{array}{cccccc}
\cos(\beta) & 0 & 0 & -\sin(\beta) & 0 & 0 \\
0 & 1 & 0 & 0 & 0 & 0 \\ 
0 & 0 & 1 & 0 & 0 & 0 \\ 
\sin(\beta) & 0 & 0 & \cos(\beta) & 0 & 0 \\ 
0 & 0 & 0 & 0 & 1 & 0 \\
0 & 0 & 0 & 0 & 0 & 1
\end{array} \right).
\end{equation*}}}
Combining these results, we obtain
\begin{equation*}
\z(\Dt_{10}, \R) \cap SO(6) = \left\{ (RZ) M(\beta)(RZ)^{-1} : M(\beta) \in \z(\overline{\D}_{10}, \R) \cap SO(6) \right\}. 
\end{equation*}
Hence, as in the tetrahedral case, the group $\z(\Dt_{10}, \R) \cap SO(6)$ is isomorphic to $S^1$ and therefore the Schur rotations associated with $\D_{10}$ are parameterised by an angle $\beta \in S^1$. As in the $\T$-case, in order to fix the boundary conditions of the transitions we consider the unique crystallographic representation $\It_{\D_{10}}$ in $B_6$ such that $\Dt_{10} = \It \cap \It_{\D_{10}}$, whose explicit form is
{\small{
\begin{equation*}
\It_{\D_{10}} = \left< \left( \begin{array}{cccccc}
0 & 0 & 0 & 0 & -1 & 0 \\
0 & 0 & 0 & 1 & 0 & 0 \\
0 & 0 & -1 & 0 & 0 & 0 \\ 
0 & 1 & 0 & 0 & 0 & 0 \\
-1 & 0 & 0 & 0 & 0 & 0 \\ 
0 & 0 & 0 & 0 & 0 & -1
\end{array} \right), \left( \begin{array}{cccccc}
0 & 0 & 0 & 0 & -1 & 0 \\ 
0 & 0 & -1 & 0 & 0 & 0 \\ 
0 & 0 & 0 & 0 & 0 & 1 \\
1 & 0 & 0 & 0 & 0 & 0 \\ 
0 & 0 & 0 & -1 & 0 & 0 \\ 
0 & -1 & 0 & 0 & 0 & 0
\end{array} \right) \right>.
\end{equation*}}}
Let $R_{\D_{10}}(\beta) := M_{\D_{10}}(\beta) R$, where $ M_{\D_{10}}(\beta) \in \z(\Dt_{10}, \R) \cap SO(6)$. We impose
\begin{equation}\label{eq_d5}
R_{\D_{10}}(\beta)^{-1} \It_{\D_{10}} R_{\D_{10}}(\beta) \simeq T_1 \oplus T_2.
\end{equation}
The corresponding system of equations (compare with \eqref{36}) has only one solution, namely $\hat{\beta} = \frac{\pi}{2}$. Hence any path $\beta_{\D_{10}}(t) : [0,1] \rightarrow S^1$ connecting $0$ with $\frac{\pi}{2}$ induces a Schur rotation as in \eqref{schur_rotation} given by $M_{\D_{10}}(\beta) \circ \beta_{\D_{10}}(t) = M_{\D_{10}}(\beta_{\D_{10}}(t)) : [0,1] \rightarrow \z(\Dt_{10},\R) \cap SO(6)$. In Figure \ref{D10_lattice} we show the quasilattice corresponding to the intermediate state $\beta = \frac{\pi}{4}$.  

\begin{table}[!t]
\begin{center}
\begin{tabular}{| c c c|}
\hline
Generator & Rep. $D_1$ & Rep. $D_2$ \\
\hline
$g_{2d}$ & $\frac{1}{2} \left( \begin{array}{ccc}
-\tau & \tau-1 & -1 \\
\tau-1 & -1 & -\tau \\
-1 & -\tau & \tau-1 
\end{array} \right)$ & $\frac{1}{2} \left( 
\begin{array}{ccc}
-1 & \tau-1 & \tau \\
\tau -1 & -\tau & 1 \\
\tau & 1 & \tau -1 
\end{array} \right)$ \\
$g_{5d}$ & $\frac{1}{2} \left( \begin{array}{ccc} 
\tau -1 & -1 & \tau \\
1 & \tau & \tau-1 \\
-\tau & \tau-1 & 1
\end{array} \right)$ & $\frac{1}{2} \left( \begin{array}{ccc}
1-\tau & -\tau & -1 \\
-\tau & 1 & 1-\tau \\
1 & \tau-1 & -\tau 
\end{array} \right)$ \\
\hline
& $P_1 = \left( \begin{array}{ccc}
0 & 1 & 0 \\
\sqrt{\frac{\tau+2}{5}} & 0 & \sqrt{\frac{3-\tau}{5}} \\
\frac{2\tau-1}{\sqrt{5(\tau+2)}} & 0 & \frac{1-2\tau}{\sqrt{5(3-\tau)}} 
\end{array} \right)$ & 
$P_2 = \left( \begin{array}{ccc}
\sqrt{\frac{3-\tau}{5}} & \frac{2\tau-1}{\sqrt{5(3-\tau)}} & 0 \\
\frac{1-2\tau}{\sqrt{5(3-\tau)}} & \sqrt{\frac{3-\tau}{5}} & 0 \\
0 & 0 & 1
\end{array} \right)$ \\
\hline
\end{tabular}
\end{center}
\caption{\label{d10_reps} Explicit forms of the representations $D_1$ and $D_2$ and the corresponding reducing matrices $P_1, P_2 \in GL(3,\R)$ (cf. \eqref{d10_reps})}
\end{table}

\subsection{Dihedral group $\D_6$}

The dihedral group $\D_6$ is isomorphic to the symmetric group $S_3$ and is generated by two elements $g_{2d}$ and $g_3$ such that $g_{2d}^2 = g_3^3 = (g_2g_3)^2 = e$. Its character table is as follows (cf. \cite{jones}):
\begin{center}
{\small{
\begin{tabular}{l|c c c }
Irrep & $E$ & $3C_2$ & $2C_3$  \\
\hline
$A_1$ & 1 & 1 & 1 \\
$A_2$ & 1 & -1 & 1 \\
$E$ & 2 & 0 & -1
\end{tabular}}}
\end{center}
In order to compute the Schur rotations associated with $\D_6$, we proceed in complete analogy with $\D_{10}$. Indeed, let $\Dt_6$ be the representation of $\D_6$ as a subgroup of $\It$ given by
{\small{
\begin{equation*}
\Dt_6 = \left< \left( \begin{array}{cccccc}
0 & 0 & 0 & 0 & 0 & -1 \\
0 & -1 & 0 & 0 & 0 & 0 \\ 
0 & 0 & 0 & 1 & 0 & 0 \\
0 & 0 & 1 & 0 & 0 & 0 \\ 
0 & 0 & 0 & 0 & -1 & 0 \\ 
-1 & 0 & 0 & 0 & 0 & 0
\end{array} \right), \left( \begin{array}{cccccc}
0 & 0 & 0 & 0 & 0 & 1 \\
0 & 0 & 0 & 1 & 0 & 0 \\ 
0 & -1 & 0 & 0 & 0 & 0 \\ 
0 & 0 & -1 & 0 & 0 & 0 \\ 
1 & 0 & 0 & 0 & 0 & 0 \\
0 & 0 & 0 & 0 & 1 & 0
\end{array} \right) \right>.
\end{equation*}}}
The matrix $R$, given by \eqref{R}, reduces this representation as 
\begin{equation}\label{D6_irreps}
\hat{\D}_6:= R^{-1} \Dt_6 R = S_1 \oplus S_2,
\end{equation}
where $S_1$ and $S_2$ are representations of $\D_6$ that are reducible. Specifically, they both split into $A_2 \oplus E$ in $GL(3,\R)$, and therefore are equivalent in $GL(3, \R)$. Using the projection operators given in \cite{fulton},  we identify two matrices  $R_1$ and $R_2$  in $GL(3,\R)$ that reduce into the same irreps $S_1$ and $S_2$, i.e. 
\begin{equation}\label{d6_reps}
R_1^{-1}S_1R_1 = R_2^{-1}S_2R_2 \simeq A_2 \oplus E.
\end{equation}
The explicit forms of such matrices are given in Table \ref{d6_reps}. Let $V$ be the matrix in $GL(6,\R)$ given by $V := R_1 \oplus R_2$. We have
\begin{equation*}
\overline{\D}_6:= V^{-1} \hat{\D}_6 V \simeq A_2 \oplus E \oplus A_2 \oplus E.
\end{equation*}
Schur's Lemma forces a matrix $P \in \z(\overline{\D}_6,\R) \cap SO(6)$ to have the form
{\small{
\begin{equation*}
P = P(\alpha,\beta) = \left( \begin{array}{cccccc}
\cos(\alpha) & 0 & 0 & -\sin(\alpha) & 0 & 0 \\
0 & \cos(\beta) & 0 & 0 & -\sin(\beta) & 0 \\
0 & 0 & \cos(\beta) & 0 & 0 & -\sin(\beta) \\
\sin(\alpha) & 0 & 0 & \cos(\alpha) & 0 & 0 \\
0 & \sin(\beta) & 0 & 0& \cos(\beta) & 0 \\
0 & 0 &  \sin(\beta) & 0 & 0 & \cos(\beta)
\end{array} \right),
\end{equation*}}}
where $(\alpha, \beta) \in S^1 \times S^1$. Hence
\begin{equation*}
\z(\Dt_6, \R) \cap SO(6) = \left\{ (RV) P(\alpha,\beta) (RV)^{-1} : P(\alpha,\beta) \in \z(\overline{\D}_6, \R) \cap SO(6) \right\}.
\end{equation*}
Therefore, contrary to the other maximal subgroups of $\I$, the Schur rotations associated with $\D_6$ are parameterised by two angles belonging to a two-dimensional torus $\mathbb{T}^2 \simeq S^1 \times S^1$. In other words, the less symmetry is preserved during the transition, the more the physical and orthogonal space are free to rotate. As before, to fix the boundary conditions, we consider the representation $\It_{\D_6}$ such that $\Dt_6 = \It \cap \It_{\D_6}$:
{\small{
\begin{equation*}
\It_{\D_6} = \left< \left( \begin{array}{cccccc}
0 & 0 & -1 & 0 & 0 & 0 \\
0 & 0 & 0 & 0 & 0 & -1 \\ 
-1 & 0 & 0 & 0 & 0 & 0 \\ 
0 & 0 & 0 & -1 & 0 & 0 \\ 
0 & 0 & 0 & 0 & -1 & 0 \\ 
0 & -1 & 0 & 0 & 0 & 0
\end{array} \right), \left( \begin{array}{cccccc}
0 & 0 & -1 & 0 & 0 & 0 \\ 
0 & 0 & 0 & 0 & 1 & 0 \\
0 & 0 & 0 & 0 & 0 & 1 \\ 
0 & -1 & 0 & 0 & 0 & 0 \\ 
0 & 0 & 0 & -1 & 0 & 0 \\ 
-1 & 0 & 0 & 0 & 0 & 0
\end{array} \right) \right>.
\end{equation*}}}
Let $R_{\D_6}(\alpha,\beta) := M_{\D_6}(\alpha,\beta) R$, where $M_{\D_6}(\alpha,\beta) \in \z(\Dt_6, \R) \cap SO(6)$. We impose
\begin{equation*}
R_{\D_6}(\alpha,\beta)^{-1} \It_{\D_6} R_{\D_6}(\alpha,\beta) \simeq T_1 \oplus T_2,
\end{equation*}
and solve for $\alpha$ and $\beta$. There are $8$ distinct solutions $(\hat{\alpha}, \hat{\beta})$ given by
{\small{ 
\begin{eqnarray*}
 (\hat{\alpha},\hat{\beta}) \in & \left\{ \left( \arctan\left(\frac{1}{2}\right), \arctan(2) \right), \left( \arctan(2), \pi -\arctan\left(\frac{1}{2}\right) \right),  \left(\arctan\left(\frac{1}{2} \right), \arctan(2)-\pi \right),  \right. \\ 
 & \left. \left(-\arctan(2), -\arctan\left(\frac{1}{2}\right)\right),   \left(\arctan\left(\frac{1}{2}\right)-\pi, \arctan(2)-\pi\right), \left( \pi-\arctan(2), -\arctan\left(\frac{1}{2}\right) \right), \right. \\   
 & \left. \left(\arctan\left(\frac{1}{2}\right)-\pi, \arctan(2)\right), \left(\pi-\arctan(2),  \pi-\arctan\left(\frac{1}{2}\right)\right) \right\}=: S_{\D_6}.
\end{eqnarray*}}}
Any path $\gamma(t) : [0,1] \rightarrow \mathbb{T}^2$ connecting $(0,0)$ with any $(\hat{\alpha}, \hat{\beta}) \in S_{\D_6}$ defines a Schur rotation $M_{\D_6}(t) := M_{\D_6}(\alpha, \beta) \circ \gamma(t) : [0,1] \rightarrow \z(\Dt_6, \R) \cap SO(6)$. 

\begin{table}[!t]
\begin{center}
\begin{tabular}{| c c c |}
\hline
Generator & Rep. $S_1$ & Rep. $S_2$ \\
\hline
$g_2$ & $\frac{1}{2} \left( \begin{array}{ccc}
-\tau & \tau-1 & -1 \\
\tau-1 & -1 & -\tau \\
-1 & -\tau & \tau-1 
\end{array} \right)$ & $\frac{1}{2} \left( 
\begin{array}{ccc}
-1 & \tau-1 & \tau \\
\tau -1 & -\tau & 1 \\
\tau & 1 & \tau -1 
\end{array} \right)$ \\
$g_3$ & $\frac{1}{2} \left( \begin{array}{ccc} 
\tau & \tau-1 & 1 \\
1-\tau & -1 & \tau \\
1 & -\tau & 1-\tau
\end{array} \right)$ & $\frac{1}{2} \left( \begin{array}{ccc}
-1 & 1-\tau & -\tau \\
\tau-1 & \tau & -1 \\
\tau & -1 & 1-\tau 
\end{array} \right)$ \\
\hline
& $R_1 = \frac{1}{\sqrt{3}} \left( \begin{array}{ccc}
\tau & 0 & 1-\tau \\
0 & \sqrt{3} & 0 \\
\tau-1 & 0 & \tau
\end{array} \right)$ & 
$R_2 = \frac{1}{\sqrt{3}} \left( \begin{array}{ccc}
0 & \sqrt{3} & 0 \\
\tau & 0 & 1-\tau \\
1-\tau & 0 & -\tau 
\end{array} \right)$ \\
\hline
\end{tabular}
\end{center}
\caption{\label{d6_reps} Explicit forms of the representations $S_1$ and $S_2$ and the corresponding reducing matrices $R_1$ and $R_2 \in GL(3,\R)$ (cf. \eqref{d6_reps})}
\end{table}

\begin{figure}[!t]
\centering
\includegraphics[scale = 0.23]{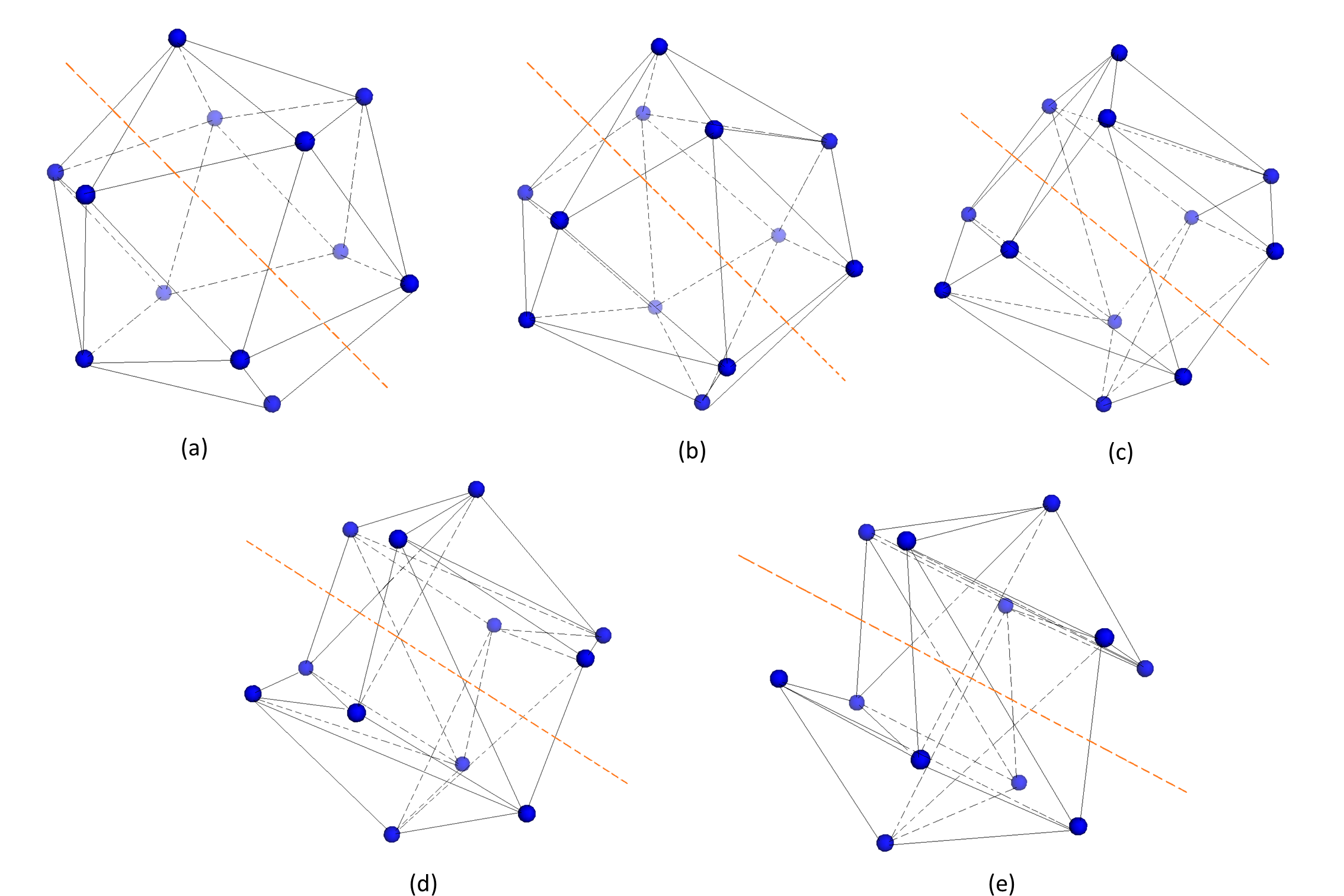}
\caption[Example of a structural transition with $\D_6$-symmetry.]{Example of a structural transition with $\D_6$-symmetry of an icosahedral point array. (a) Initial configuration for $t=0$, corresponding to the vertices of an icosahedron. (b) Resulting array for $t = 0.25$. (c) The intermediate point set ($t = 0.5$), forming the vertices of a hexagonal prism. (d) Transformed array for $t = 0.75$. (e) Final state of the transition ($t =1$): the point set forms the vertices of an icosahedron, albeit different from the initial one. The dashed red line corresponds to a three-fold axis of the arrays that is fixed during the entire transition. The lines indicate the relative positions of the icosahedral vertices during the transition. }
\label{D6_fig}
\end{figure}

\begin{figure}[!t]
\centering
\includegraphics[scale = 0.13]{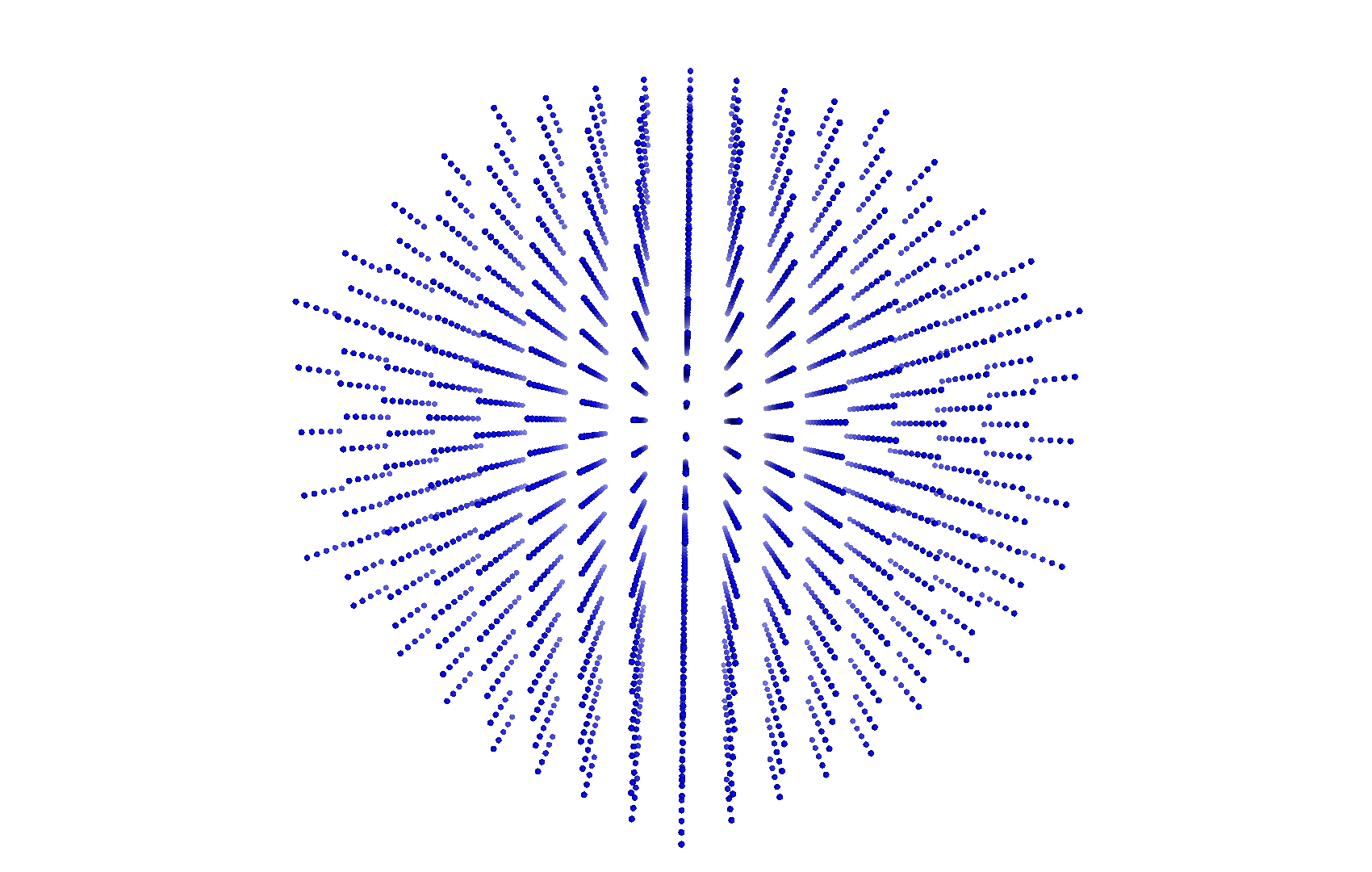}
\caption[A three-dimensional lattice obtained from a transition with $\D_6$-symmetry.]{The three-dimensional lattice obtained from the transition analysed in Figure \ref{D6_fig} and corresponding to the intermediate state $t = 0.5$.  }
\label{D6_lattice}
\end{figure}

\paragraph{Continuous transformation of an icosahedron into a hexagonal prism.} Let us consider the path $\gamma : [0,1] \rightarrow \mathbb{T}^2$ given by $\gamma(t) = (t\hat{\alpha}, t\hat{\beta})$, connecting $(0,0)$ with the point  $(\hat{\alpha}, \hat{\beta}) = \left( \arctan\left(\frac{1}{2}\right), \arctan(2) \right) \in S_{\D_6}$, and let $M_{\D_6}(t)$ be the corresponding Schur rotation. We consider the point sets $\C_0$ given by the projection into $E^{\parallel}$ of the orbit under $\widetilde{\I}$ of the lattice point $\bfe_1 = (1,0,0,0,0,0)$:
\begin{equation*}
\C_0 := \pi^{\parallel}\left(\mathcal{O}_{\It}(\bfe_1)\right) = \left\{ \pi^{\parallel}(A\bfe_1) : A \in \It \right\}.
\end{equation*}
The points of $\C_0$ constitute the vertices of an icosahedron (see Figure \ref{D6_fig} (a)). The Schur rotation $M_{\D_6}(t)$ induces a continuous transformation of $\C_0$ via the corresponding projection operators $\pi^{\parallel}_t$; in particular, we consider the family of finite point set $\C_t$ as in \eqref{clusters}. In Figure \ref{D6_fig} we plot these point sets for $t =0$, $0.25$, $0.5$, $0.75$ and $1$: we notice that the icosahedron ($t = 0$) is continuously transformed into a hexagonal prism ($t = 0.5$), and the array for $t = 1$ forms the vertices of an icosahedron, that is distinct from the initial one. The three-fold axis highlighted is fixed during the transition, and the point sets $\C_t$ are invariant under the action of the representation $S_1$ of $\D_6$ as in the decomposition \eqref{D6_irreps}. In analogy to the case of the tetrahedral transition, the corresponding model set for $t = 0.5$ defines a lattice in $E^{\parallel}$ (see Figure \ref{D6_lattice}).   

\section{Conclusions}\label{conclusions}

In this paper we characterised structural transitions of icosahedral structures (both infinite and finite) in a group theoretical framework. We generalised the Schur rotation method, proposed in \cite{kramer, kramer2}, for transitions with $\G$-symmetry, where $\G$ is a maximal subgroup of the icosahedral group. This allowed for the identification of the order parameters of the transitions, which correspond to the possible angle(s) of rotation of the physical and orthogonal spaces that are embedded into the higher dimensional space. 
In this paper, we have considered embeddings into 6D, which corresponds to the minimal embedding dimension in which the icosahedral group is crystallographic. For this case, we have characterised transitions exhaustively here. However, there is the possibility to embed the symmetry groups into even higher dimensions, which would open up additional degrees of freedom. This could be important for those transitions which in 3D projection result in sigularities as lattice vectors in projection collapse upon each other during the transition. An example are the two opposing vertices of the symmetry axis defining $D_{10}$ symmetry, which would meet at the origins during the $D_{10}$-preserving transition identified here. The additional degrees of freedom could be used to construct transition paths avoiding this. 

With the mathematical formalism developed here, we were able to easily prove the relation between our approach and the Bain strain method in \cite{paolo}.  Moreover, based on the subgroup structure analysis carried out in \cite{zappa1}, we classified the boundary conditions of the transitions, which correspond to angles of a $k$-dimensional torus, and provided specific examples of such transitions for icosahedral model sets and point arrays.   

This work paves the way for a dynamical analysis of such transitions. In particular, Ginzburg-Landau energy functions can be formulated, which depend on the order parameters of the transition and account for the symmetry breaking of the system, in line with previous models \cite{zappa}. The symmetry arguments developed here should give a better understanding of the energy landscapes, with (local) minima corresponding to phases with higher symmetry, and possibly of the transition pathways between them, which can be identified as paths on a torus connecting two icosahedral boundaries. 

As mentioned in the Introduction, icosahedral finite point sets play an important role in the modeling of material boundaries of viral capsids. A complete theoretical framework of conformational changes of viral capsids is still lacking; previous work \cite{giuliana} analysed such structural transformations with the Bain strain approach, by embedding the point arrays constituting the blueprints for the capsid into six-dimensional lattices and then considering possible transformations of these higher dimensional point sets. In a similar way, the results developed here can provide new insights into the dynamics of conformational changes in viruses, and a better understanding of their maturation processes. Finally, point arrays with icosahedral symmetry have been used in the modeling of fullerenes and carbon onions \cite{twarock1, dechant1}, and therefore, the new mathematical models are likely to have wider applications in science.    

\paragraph{Acknowledgements} We thank David Salthouse, Paolo Cermelli and Giuliana Indelicato for useful discussions. RT gratefully acknowledges a Royal Society Leverhulme Trust Senior Research fellowship (LT130088) and ECD a Leverhulme Trust Early Career Fellowship (ECF2013-019). 

\bibliographystyle{unsrt}
\bibliography{transitions_biblio}

\end{document}